\documentclass[12pt,preprint]{aastex}

\begin{document}

\title{Primordial Stellar Feedback and the Origin of \\
Hyper Metal-poor Stars}
\author{Torgny Karlsson}
\affil{NORDITA, Blegdamsvej 17, DK-2100 Copenhagen {\O}, Denmark}
\email{karlsson@nordita.dk}
\shorttitle{The Origin of Hyper Metal-poor Stars}
\shortauthors{Karlsson}
\begin{abstract}
\noindent
The apparent absence of stars in the Milky Way halo with \mbox{$-5\lesssim[\mathrm{Fe}/\mathrm{H}]\lesssim-4$} suggests that the gas out of which the halo stars were born experienced a period of low or delayed star formation after the local universe was lit up by the first, metal-free generation of stars (Pop~III). Negative feedback owed to the Pop~III stars could initially have prevented the pre-Galactic halo from cooling, which thereby delayed the collapse and inhibited further star formation. During this period, however, the nucleosynthesis products of the first supernovae (SNe) had time to mix with the halo gas. As a result, the initially primordial gas was already weakly enriched in heavy elements, in particular iron, at the time of formation of the Galactic halo. The very high, observed $\mathrm{C}/\mathrm{Fe}$ ratios in the two recently discovered hyper metal-poor stars ($[\mathrm{Fe}/\mathrm{H}]<-5$) \mbox{HE 0107$-$5240} and \mbox{HE 1327$-$2326} as well as the diversity of $\mathrm{C}/\mathrm{Fe}$ ratios in the population of extremely metal-poor stars ($[\mathrm{Fe}/\mathrm{H}]<-3$) are then naturally explained by a combination of pre-enrichment by Pop~III stars and local enrichment by subsequent generations of massive, rotating stars, for which the most massive ones end their lives as black hole-forming SNe, only ejecting their outer (carbon-rich) layers. The possible existence of populations of mega metal-poor/iron-free stars ($[\mathrm{Fe}/\mathrm{H}]<-6$) is also discussed.

\end{abstract}
\keywords{Stars: abundances -- Stars: individual (HE 0107$-$5240, HE 1327$-$2326) -- Stars: Population II -- Galaxy: evolution -- Galaxy: formation}

\section{Introduction} \label{intro}
\noindent
The unique chemical abundance patterns imprinted in the old, extremely metal-poor ($[\mathrm{Fe}/\mathrm{H}]\footnote{$[\mathrm{A}/\mathrm{B}] \equiv \log(n_{\mathrm{A}}/n_{\mathrm{B}}) - \log(n_{\mathrm{A}}/n_{\mathrm{B}})_{\odot}$, where $n_{\mathrm{A}}$ and $n_{\mathrm{B}}$ are the number densities of element A and B, respectively, and where $\log(n_{\mathrm{A}}/n_{\mathrm{B}})_{\odot}$ denotes the logarithmic abundance ratio of A and B in the Sun.}<-3$, Beers \& Christlieb 2005\nocite{bc05}), stellar population of the Milky Way's halo play a key role in understanding the formation of the Galaxy and the nature of the first stars. During the formation epoch of this extremely metal-poor population, the presence of carbon, as well as oxygen, was crucial for low-mass star formation to occur as these elements act as efficient cooling agents in interstellar gas (Bromm \& Loeb 2003\nocite{bromm03}). A curiously large star-to-star scatter is also found in, e.g., $\mathrm{C}/\mathrm{Fe}$ for the stars in this metallicity regime. Two stars stick out in particular; \mbox{HE 0107$-$5240} (Christlieb et al. 2002\nocite{cetal02}) and \mbox{HE 1327$-$2326} (Frebel et al. 2005\nocite{frebel05}). They show the hitherto highest reported enhancements of carbon of $[\mathrm{C}/\mathrm{Fe}] \sim 4$ and are both depleted in iron a factor of about $20$ with respect to the next most iron-poor stars at $[\mathrm{Fe}/\mathrm{H}]=-4$. As such, they are defined as carbon-enhanced ($[\mathrm{C}/\mathrm{Fe}]>1$), hyper metal-poor stars ($[\mathrm{Fe}/\mathrm{H}]<-5$, Beers \& Christlieb 2005\nocite{bc05}).

Several scenarios trying to explain the individual chemical abundance patterns of \mbox{HE 0107$-$5240} and \mbox{HE 1327$-$2326} have been discussed in the literature including accretion of interstellar matter, accretion of processed material from a binary companion, self-enrichment, and pre-enrichment of the molecular clouds out of which the stars were born (see, e.g., Beers \& Christlieb 2005\nocite{bc05} and references therein). In particular, the black hole-forming supernova (SN) models by Umeda \& Nomoto (2003\nocite{un03}) can successfully fit most of the relative abundance ratios in both \mbox{HE 0107$-$5240} and \mbox{HE 1327$-$2326} (Iwamoto et al. 2005\nocite{iwamoto05}). This, however, requires fine-tuning of the parameters controlling the amount of mixing and fall-back in the models. More importantly, although the metal-poor tail of the observed Galactic halo metallicity distribution suffers from small number statistics, the presence of stars below \mbox{$[\mathrm{Fe}/\mathrm{H}]=-5$}, or rather, the apparent lack of stars in the range \mbox{$-5\lesssim [\mathrm{Fe}/\mathrm{H}]\lesssim -4$} is intriguing and cannot be explained by the models of Umeda \& Nomoto (2003\nocite{un03}).

Below, we will explore an alternative formation scenario for the hyper metal-poor stars and argue that star formation was suppressed during the initial phase of the collapse of the Galactic halo. Such a suppression would, as we shall see, lead to a gap in the halo metallicity distribution and retain a large scatter in $\mathrm{C}/\mathrm{Fe}$, even for unevolved, single, extremely metal-poor stars.

\section{Star Formation and Chemical Evolution in the Early Galaxy}
\noindent   
Star formation in the pre-Galactic halo could have been inhibited by negative feedback mechanisms by the first (Pop~III) stars. Lyman-Werner radiation (dissociating $\mathrm{H}_2$), ultra-violet radiation (ionizing $\mathrm{H}$), and Pop~III SN explosions (heating and removing gas) could initially have prevented the halo from cooling, collapsing, and forming new stars (see e.g., Reed et al. 2005\nocite{reed05} for a discussion on feedback effects). Such an early feedback phase has also been proposed as a solution to the so called angular momentum problem in simulations of disk galaxy formation (e.g., Sommer-Larsen, G\"{o}tz, \& Portinari 2003\nocite{sl03}). Hence, when the collapse of larger structures eventually occurred and the gas became re-accessible to star formation, the Galactic halo had already been enriched in metals, e.g., iron, by the first generation of SNe. Stars formed in the beginning of the second star formation epoch may therefore have contained small traces of iron, to a level of \mbox{$[\mathrm{Fe}/\mathrm{H}]<-5$}. Concurrently, ejecta from the second generation of SNe were dispersed in the interstellar medium (ISM) of the halo. However, due to the increased star formation rate (SFR), this material did not have time to mix as much with its surroundings before locked up in subsequent generations of stars. Later stellar generations should therefore display significantly higher abundances of iron, i.e., \mbox{$[\mathrm{Fe}/\mathrm{H}]\gtrsim -4$}. We shall now quantify this idea.

The stochastic chemical enrichment model used in the present study is described elsewhere (Karlsson 2005\nocite{karlsson05}; Karlsson \& Gustafsson 2005\nocite{kg05}). Here we will only briefly mention the main features of the model including new improvements. Since observational data on oxygen are uncertain (Garc\'{i}a-p\'{e}rez et al. 2006\nocite{aegp06}) and generally less available for stars below $[\mathrm{Fe}/\mathrm{H}]=-3$, we will, apart from iron, focus on the evolution of carbon.

The ejecta of each exploding star are assumed to be evenly spread within a continuously growing volume $V_{\mathrm{mix}}(t)$, the mixing volume. The mixing volume growth is caused, e.g., by turbulent motions and cloud-cloud collisions in the ISM and is modelled as a random-walk process. Hence, \mbox{$V_{\mathrm{mix}}(t)=4\pi\times (\sigma_{\mathrm{mix}} t)^{3/2}/3$}, where the diffusion coefficient is set to \mbox{$\sigma_{\mathrm{mix}}=1.12\times 10^{-3}$ kpc$^{2}$ Myr$^{-1}$}, within the limits estimated from data given in Bateman \& Larson (1993\nocite{bl93}). Star formation is assumed to be spatially uncorrelated, i.e., $\psi=\psi(t)$, implying that exploding stars are randomly distributed in space. With this assumption, the probability of finding a region enriched by the ejecta of $k$ stars is given by the Poisson distribution \mbox{$w_{\mathrm{ISM}}(k,t)=e^{-\mu(t)}\mu(t)^k/k!$}, where $\mu(t)$ is the average number of stars contributing to the enrichment in a random point in space at time $t$.

\begin{figure}[t]
\resizebox{\hsize}{!}{\includegraphics{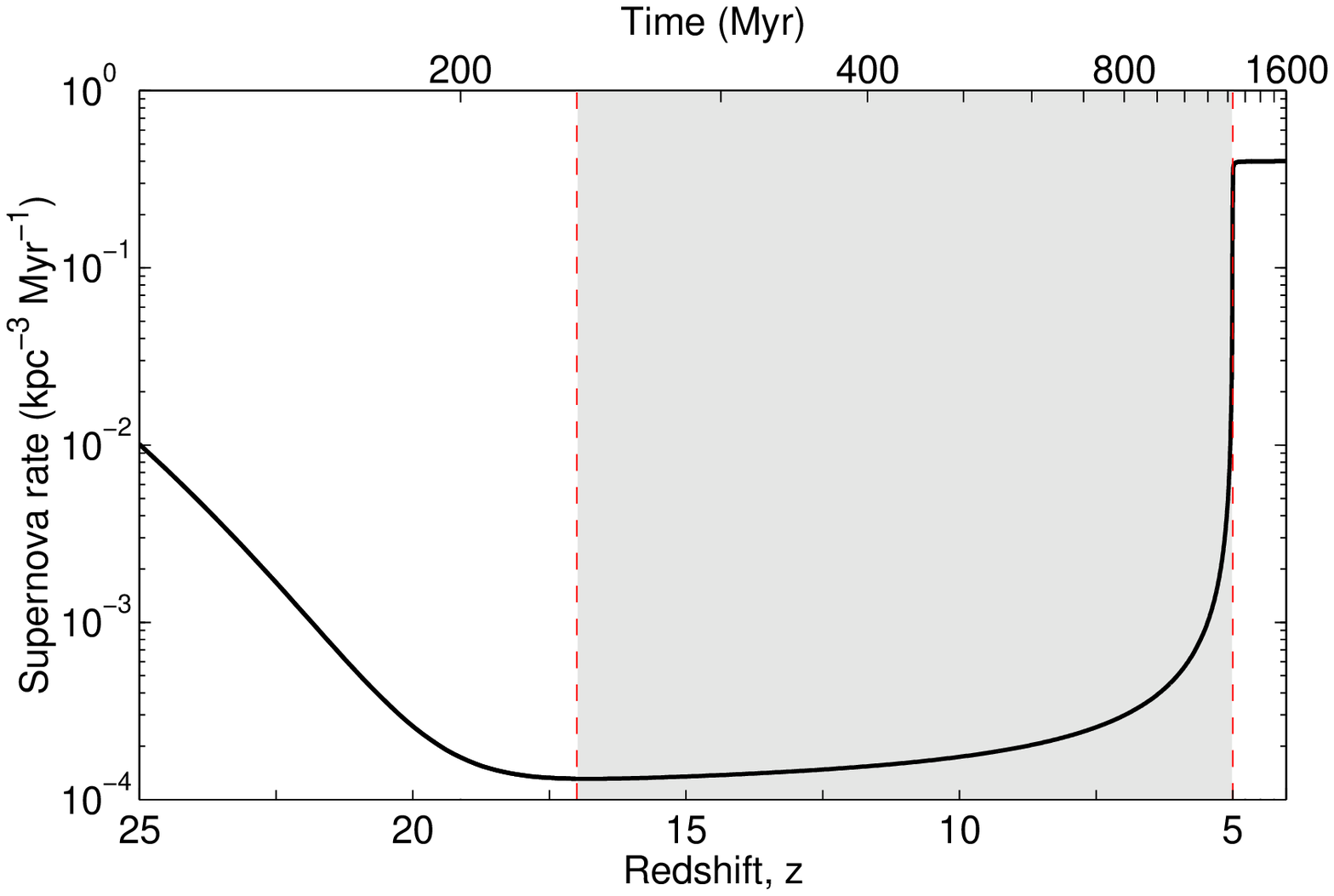}}
\caption{The adopted SFR in the Galactic halo (black line) as given by the fraction of massive stars exploding as SNe. The shaded area indicates the period of low SFR.}
\label{sfr}
\end{figure}

In order to properly model the build-up of carbon in the ISM, the contribution from intermediate mass stars should be included, taking into account the delay due to non-zero stellar lifetimes. Hence, the SN rate $u_{\mathrm{SN}}$ should, in all relevant equations, be replaced by the "stellar enrichment rate" $u_{\mathrm{SER}}$, defined as the total stellar death rate at time $t$. The death rate of stars of mass $m$, $u_m$, is given by  

\begin{equation}
u_{m}(t)=
\left\{ \begin{array}{ll}
\!\!0,\!\! & \!\!t<g_{\tau}(m_u)\!\\
\!\!\phi(m)\!\times \!\psi(t-g_{\tau}(m)),\!\! & \!\!t\ge g_{\tau}(m_u),\!
\end{array}
\right.
\label{um}
\end{equation}

\noindent
where $\phi(m)\propto m^{-2.3}$ (Kroupa 2001\nocite{kroupa01}) is the initial mass function (IMF), $\psi(t)$ is the SFR (Figure \ref{sfr}), and $g_{\tau}(m)$ is the lifetime of a star of mass $m$. Hence, 

\begin{equation}
u_{\mathrm{SER}}(t)=\int\limits_{g_{\tau}^{-1}(t)}^{m_u}u_m(t)\mathrm{d}m,
\label{user}
\end{equation}

\noindent
for $t\ge g_{\tau}(m_u)$ and zero otherwise. The lower mass limit is determined by the least massive star able to enrich the ISM at time $t$ while the upper mass limit $m_u=60~\mathcal{M_{\odot}}$ is fixed.

Relaxing the instantaneous recycling approximation we need to calculate a mixing mass distribution (for details, see Karlsson 2005\nocite{karlsson05}) for each stellar mass $m$, i.e., 

\begin{eqnarray}
f_{\mathcal{M},M}^{k} & = & c'_k\int_{\Delta D_{M}}a_{\star}(t)V_{\mathrm{mix}}(\tau_V)w_{\mathrm{ISM}}(k-1,t) \nonumber\\
& \times & u_{m}(t-\tau_V)\psi(t)\mathrm{d}t\mathrm{d}\tau_V,
\label{fmmk}   
\end{eqnarray}

\noindent
where the total supernova rate $u_{\mathrm{SN}}$ is replaced by $u_{m}$. The function $a_{\star}(t)$ denotes the fraction of still-surviving stars formed at time $t$ and $c_k'$ is a normalization factor. The above probability density function $f_{\mathcal{M},M}^{k}$ is a generalization of the mixing mass distribution $f_{M_k}$ derived in Karlsson (2005\nocite{karlsson05}) and simultaneously gives the stellar masses (i.e. the yields) {\it and} the mixing masses (i.e., the amount of dilution of the ejecta) of $k$ enriching stars.

As an efficient cooling agent, carbon presumably played an active role in the formation of extremely metal-poor stars by promoting fragmentation of interstellar gas clouds. This was accounted for by suppressing low-mass star formation in gas with $[\mathrm{C}/\mathrm{H}]<-3.5$ (Bromm \& Loeb 2003\nocite{bromm03}). Furthermore, since SN-induced star formation partly occurs in the ISM, this was introduced by increasing the the probability of forming low-mass stars in close connection to SN remnants. A minimum mass of $M_{\mathrm{mix}}=3\times 10^4~\mathcal{M_{\odot}}$ was, however, assumed to be swept up by the remnants before stars were allowed to form (Ryan, Norris, \& Beers 1996\nocite{rnb96}). The inclusion of SN-induced star formation is not crucial for the conclusions presented here.

Carbon yields by Meynet \& Maeder (2002\nocite{mm02}) derived from rotating intermediate- and high-mass star models (metallicity $Z=10^{-5}$) were used. Iron yields for SNe in the mass range $13\le m/\mathcal{M_{\odot}}\le 30$ ($Z=0$) were taken from Umeda \& Nomoto (2002\nocite{un02}). SNe with extensive fall-back, i.e., black hole-forming SNe, were assumed to form in the mass range $30 < m/\mathcal{M_{\odot}}\le 60$ and their iron yields were set to zero. The small amount of iron supposedly produced in O-Ne-Mg core-collapse SNe ($8\lesssim m/\mathcal{M_{\odot}}\lesssim 10$) was, however, taken into account (Mayle \& Wilson 1988\nocite{mw88}). Yields of pair-instability SNe were not included (however, see \S~\ref{discon}).

The effect of the primordial stellar feedback was realized by assuming an exponential decay of the local (Pop~III) star burst to a low value (cf. Sommer-Larsen et al. 2003\nocite{sl03}), followed by an increase of the SFR to a value of $0.4$ kpc$^{-3}$ Myr$^{-1}$, appropriate for the Galactic halo during this period of time (Samland, Hensler, \& Theis 1997\nocite{samland97}). The star burst was normalized to $0.12$ SNe per kpc$^3$ over $100$ million years ($25\ge z\ge 17$). This particular value is arbitrary but lies in between different estimates of the early cosmic SFR (e.g., Springel \& Hernquist 2003\nocite{sphe03}; Daigne et al. 2004\nocite{daigne04}). The adopted SFR is shown in Figure \ref{sfr}. The dashed line at $z=17$~($\pm 4,~68\%$ confidence errors) marks the redshift of reionization as suggested by the WMAP data (Spergel et al. 2003\nocite{spergel03}) while the line at $z=5$ marks the adopted redshift of formation of the Milky Way galaxy. These redshifts were used to estimate, respectively, the end of the first star formation epoch (Pop~III) and the beginning of the second, main star formation epoch in the Galactic halo. As we will see below, an initial star burst followed by a quiescent period of low star formation will generate a "pre-enrichment"-peak in the halo metallicity distribution. Such a peak would be absent for non-bimodal (e.g., constant or increasing/decreasing) SFRs (see Karlsson 2005, Figure 6). The particle density of the ISM was set to $0.1$ cm$^{-3}$.

\section{Results}\label{results}
\noindent
Figure \ref{fz} shows the predicted halo metallicity distribution. Owed to the bimodality of the SFR, a small group of hyper metal-poor stars appears in the range \mbox{$-5.5\lesssim [\mathrm{Fe}/\mathrm{H}]\lesssim -5$} (implying a mixing mass of $M_{\mathrm{mix}}\sim 10^7~\mathcal{M_{\odot}}$). These stars were formed out of pre-enriched gas in the beginning of the second star formation epoch at $z=5$ (cf. Umeda \& Nomoto 2003\nocite{un03}). Depending on whether the ISM at that point was, on average, enriched by $1$ SN or many SNe, stars in this group would either be found in a few narrow peaks (the present model) or in one broader peak. The predicted number of stars at $[\mathrm{Fe}/\mathrm{H}]<-5$ is in good agreement with observations only if suppression of low-mass star formation below $[\mathrm{C}/\mathrm{H}]=-3.5$ is taken into account (Bromm \& Loeb 2003\nocite{bromm03}). This suggests that the primordial/early IMF differed from that of the present-day, otherwise hyper metal-poor stars would have been more abundant.

The predicted distribution of stars in the $[\mathrm{C}/\mathrm{Fe}]-[\mathrm{Fe}/\mathrm{H}]$ plane is shown in Figure \ref{cfeh}. The star-to-star scatter is large and a significant fraction of the model stars display a carbon enhancement in excess of $[\mathrm{C}/\mathrm{Fe}]=1$. Since low-mass stars presumably were unable to form in gas with $[\mathrm{C}/\mathrm{H}]\lesssim -3.5$, hyper metal-poor stars are expected to be strongly carbon-enhanced, as observed. The absence of stars with iron abundances $-5\lesssim [\mathrm{Fe}/\mathrm{H}]\lesssim -4$ is seen as a depression in the density function. The size of this gap depends primarily on the SFR, as well as on $V_{\mathrm{mix}}$, cf. Eq. (\ref{fmmk}), while the number of (carbon-enhanced) stars in the peak(s) below $[\mathrm{Fe}/\mathrm{H}]=-5$ also depends on the fraction of black hole-forming SNe. Note that, irrespectively of iron abundance, the majority of low-mass stars below $[\mathrm{Fe}/\mathrm{H}]=-3$ were formed during the initial phase of the second star formation epoch.

In the present model, nearly $100\%$ of the ISM was affected by ejecta from intermediate-mass stars at \mbox{$z=5$}. However, the amount of carbon produced in this type of stars is not nearly enough to enrich the ISM to the level required by observations. The highest abundances of carbon can only be found near recent explosion sites of massive, black hole-forming SNe. These stars produce a lot of carbon and if all of it escapes the black hole, the ambient medium should indeed exhibit high $\mathrm{C}/\mathrm{H}$ ratios. Nonetheless, for the generic set of yields (Meynet \& Maeder 2002, Figure \ref{cfeh}a), the model cannot explain the presence of the most carbon-rich stars. As suggested in Figure \ref{cfeh}b, this discrepancy could largely be due to uncertainties in the calculations of the stellar yields. Using a set of models with somewhat different input physics, Meynet, Ekstr{\"o}m, \& Maeder (2006\nocite{mem05}) found that the carbon yield increased a factor of $4$ with respect to their generic calculations, at least for the most massive stars. It should also be noted that 1D metal-poor model atmospheres overestimate carbon abundances derived from molecular lines with as much as $0.6-0.7$ dex (see Collet, Asplund, \& Trampedach 2005\nocite{collet05}; Frebel et al. 2006\nocite{frebel06}).

\nocite{beers99}

\nocite{barklem05}
\nocite{spite05}
\nocite{plez05}
\nocite{aoki04}
\nocite{aoki02}
\nocite{cetal02}
\nocite{frebel05}

The scenario presented here predicts the possible existence of a small group of mega metal-poor stars ($[\mathrm{Fe}/\mathrm{H}]<-6$, see Figure \ref{cfeh}). These stars would be enriched by a Pop~III O-Ne-Mg SN (plus $\ge 1$ second generation black hole-forming SN). If so, as many as one out of five hyper metal-poor stars could be a mega metal-poor star. Their precise location in the $[\mathrm{C}/\mathrm{Fe}]-[\mathrm{Fe}/\mathrm{H}]$ diagram is, however, uncertain due to the uncertain Fe yield of O-Ne-Mg SNe. Not visible in Figure \ref{cfeh} is a second group of stars. These stars are formed out of gas only enriched by material from massive, second generation black hole-forming SNe. Although they are effectively iron-free, they may have a carbon abundance as high as $[\mathrm{C}/\mathrm{H}]\sim -1.5$, similar to the mega and hyper metal-poor stars. The predicted fraction of Fe-free stars in the Galactic halo is estimated to $\sim 10^{-4}$. This fraction, however, depends sensitively on the SFR and the adopted primordial IMF (e.g., Nakamura \& Umemura 2001\nocite{nu01}; Abel, Bryan, \& Norman 2002\nocite{abel02}).

\section{Discussion and Conclusions}\label{discon}
\noindent
An alternative source to the pre-enrichment of the Galactic halo may be the elusive pair-instability SNe, whose very massive progenitors ($140\lesssim \mathcal{M}/\mathcal{M_{\odot}}\lesssim 260$) have been suggested to dominate the first generation of stars (Bromm, Coppi, \& Larson 1999\nocite{bromm99}; Abel et al. 2002\nocite{abel02}; but see, e.g., Silk \& Langer 2006\nocite{sl05}). If very massive stars indeed were formed in collapsing mini-halos surrounding the Galaxy, the reason why we have yet not found any trace of their nucleosynthetic signature in stars above $[\mathrm{Fe}/\mathrm{H}]=-4$ (see e.g., Tumlinson, Venkatesan, \& Shull 2004\nocite{tvs04}) could thus be due to the fact that their ejecta were highly diluted and mixed with much less diluted ejecta of normal core-collapse SNe before locked up in low-mass stars. If so, {\it the only observable signature of pair-instability SNe would be imprinted in the iron-peak of hyper metal-poor stars}. It is, however, not clear whether their ejecta were able to mix with enough mass of interstellar gas ($\sim 10^9~\mathcal{M_{\odot}}$) prior to the collapse of the Galactic halo.

Apart from carbon, strong enhancements of nitrogen and oxygen are also measured in hyper metal-poor stars. For example, the $[\mathrm{N}/\mathrm{C}]$ ratio in HE 0107$-$5240 and HE 1327$-$2326 (subgiant solution) are $-1.1$ and $+0.3$, respectively (Frebel et al. 2006\nocite{frebel06} and references therein). Non-rotating models of massive stars produce very little primary nitrogen (via the CNO-cycle) since freshly synthesized $\mathrm{C}$ and $\mathrm{O}$ never reach the $\mathrm{H}$-burning shell (see, e.g., Umeda \& Nomoto 2002\nocite{un02}; Meynet et al. 2006\nocite{mem05}). In contrast, rotating (metal-poor) models are able to produce large amounts of nitrogen and indeed, the set of models (with different $v_{\mathrm{ini}}$ and $Z$) by Meynet et al. (2006\nocite{mem05}) encompasses the observed range in $[\mathrm{N}/\mathrm{C}]$. Rotating massive stars thus provide a plausible explanation to the enhancements of nitrogen in hyper metal-poor stars, in agreement with the above hypothesis. However, additional enrichment, such as mass-transfer from a possible binary companion, cannot be ruled out at this point. Regarding oxygen, a fraction of the hyper metal-poor stars in the simulations tend to be too oxygen-rich as compared to present observations. Since the oxygen yield depends strongly on stellar mass, the discrepancy can be reduced by assuming a lower limiting mass $m_u$ for direct black hole-formation. However, a number of extremely metal-poor stars are displaying oxygen abundances in excess of \mbox{$[\mathrm{O}/\mathrm{H}]=-2$} (e.g., CS 29498$-$43, Aoki et al. 2004), in accordance with predictions. Thus, there may be hyper metal-poor stars with $[\mathrm{O}/\mathrm{H}]>-2$ yet to be found. This merits further investigation.

\begin{figure}[t]
\resizebox{\hsize}{!}{\includegraphics{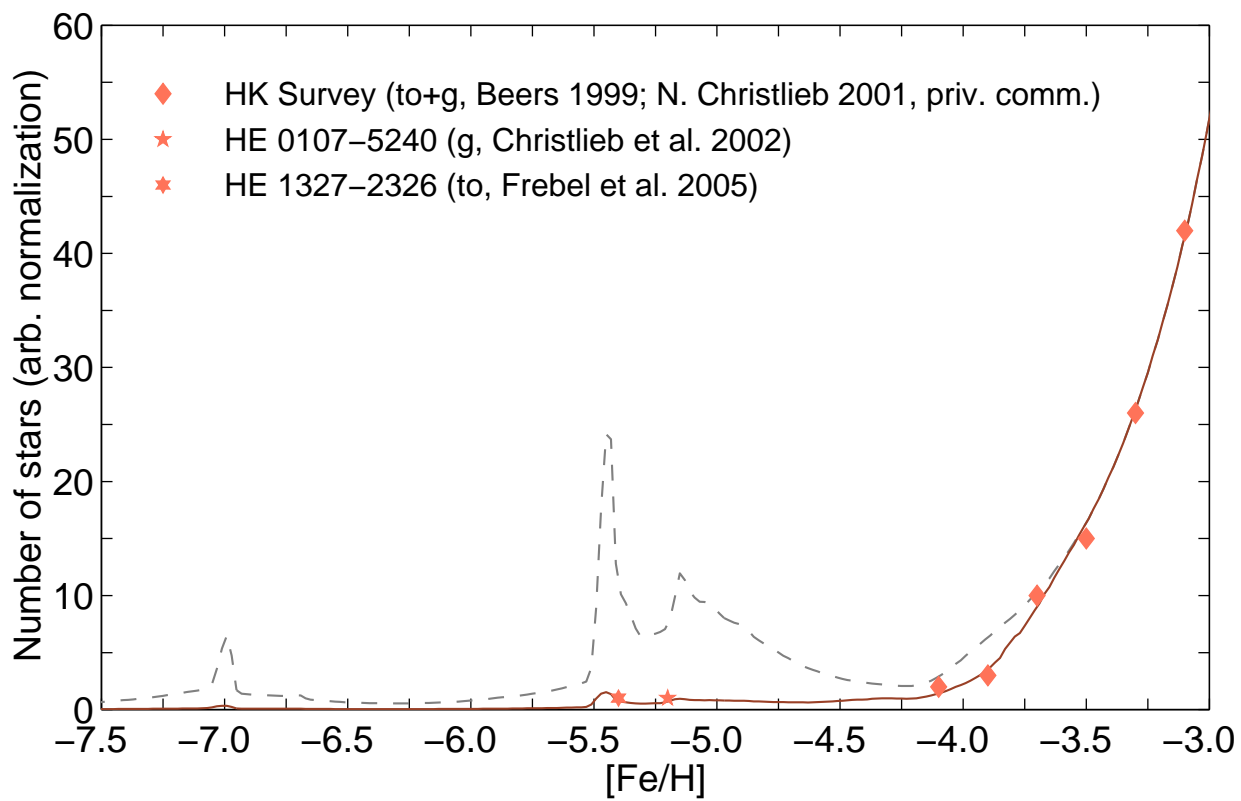}}
\caption{The predicted halo metallicity distribution. The dashed line denotes the distribution of stars neglecting the suppression of low-mass star formation in carbon-deficient gas. The full line denotes the fraction of stars with carbon abundance $[\mathrm{C}/\mathrm{H}]\ge -3.5$. In both cases, a distinct population of stars appears in the range $-5.5 \lesssim [\mathrm{Fe}/\mathrm{H}] \lesssim -5$, in agreement with observations (symbols).}
\label{fz}
\end{figure}

\begin{figure}[t]
\resizebox{\hsize}{!}{\includegraphics{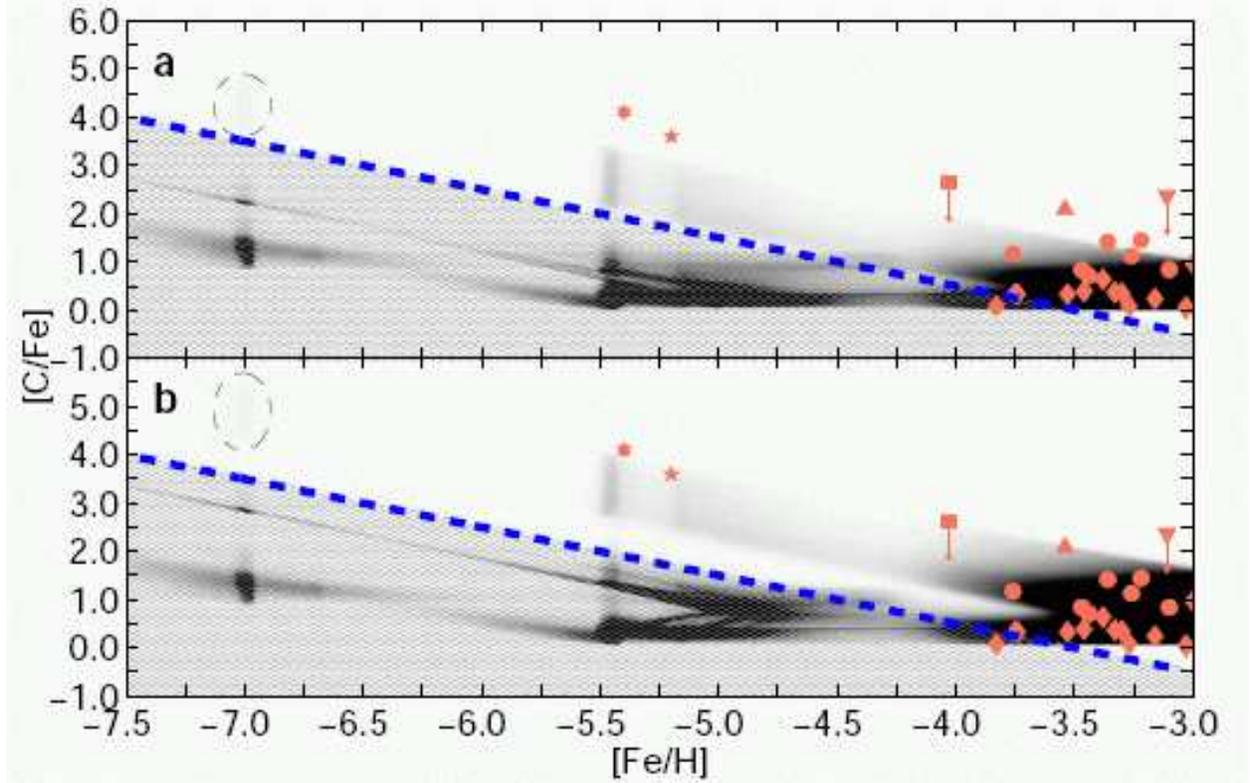}}
\caption{The predicted distribution of stars in the $[\mathrm{C}/\mathrm{Fe}]-[\mathrm{Fe}/\mathrm{H}]$ plane. {\it Upper panel:} The thick, dashed line indicates a carbon abundance of $[\mathrm{C}/\mathrm{H}]=-3.5$. The shaded area below this limit should contain no low-mass stars (cf. Figure 2). The symbols denote observations of stars in the Milky way's halo. The circles are unevolved dwarf/subgiant stars (Barklem et al. 2005), the diamonds are stars on the lower red giant branch (Spite et al. 2005), the square is \mbox{G 77$-$61} (Plez \& Cohen 2005), the upward facing triangle is \mbox{CS 29498$-$43} (Aoki et al. 2004), the downward facing triangle is \mbox{CS 22957$-$027} (Aoki et al. 2002), and finally, the pentagon and hexagon are \mbox{HE 0107$-$5240} (Christlieb et al. 2002) and \mbox{HE 1327$-$2326} (Frebel et al. 2005), respectively. \mbox{G 77$-$61} and \mbox{CS 22957$-$027} are known to be members of binary systems and may have been born with a lower surface carbon abundance as indicated by the arrows. The predicted group of mega metal-poor stars is encircled. {\it Bottom panel:} same as above but with a carbon yield increased by a factor of $4$ for stars in the mass range $30 \le m/\mathcal{M_{\odot}} \le 60$. Note that observed abundance ratios are not corrected for 3D effects.}
\label{cfeh}
\end{figure}

Observations of extremely and hyper metal-poor stars provide valuable information on the conditions at the time of formation of the Galaxy and it is imperative to, e.g., establish the metallicity distribution in these regimes. As discussed above, the lack of halo stars in the range $-5\lesssim [\mathrm{Fe}/\mathrm{H}]\lesssim -4$ (Figure \ref{fz}) suggests that the Galactic halo experienced a period of low star formation (star formation might have still occurred in e.g., the Galactic bulge) posterior to a prompt injection of metals by the first generation of stars. Moreover, the very strong enhancements of carbon, as well as of nitrogen and oxygen, in \mbox{HE 0107$-$5240} and \mbox{HE 1327$-$2326} are comparable to what is expected from subsequent enrichment of the ISM by rotating, black hole-forming SNe alone, without invoking additional sources (Figure \ref{cfeh}). A significant fraction of the extremely metal-poor stars is also expected to be carbon-enhanced. Finally, if the inner and outer halo collapsed at different times, a correlation between iron abundance and orbital eccentricity of the hyper metal-poor stars may be expected to exist. If observed, such a correlation can put strong constraints on the formation of the Milky Way halo. Further modelling is, however, necessary in order to determine the level of pre-enrichment and estimate effects of violent merging and density gradients in the ISM.

\begin{acknowledgements}
\noindent
I thank G. Meynet for discussions and for giving me a preprint of Meynet et al. (2006)\nocite{mem05} in advance. I also thank A. Andersen, R. Collet, M. Davies, A. Frebel, B. Gustafsson, and E. Olsson, as well as the anonymous referee, for valuable comments and suggestions.

\end{acknowledgements}

\end{document}